\documentstyle[preprint,aps]{revtex}

\begin{document}
\preprint{SNUTP/98-065, MIT-CTP/2766, gr-qc/9807083}

\title{Black Hole Entropy and Exclusion Statistics}
\author{Hyeong-Chan Kim${}^{{\rm a}}$\thanks{E-mail: 
leo@newton.skku.ac.kr},
Yoonbai Kim${}^{{\rm a}}$\thanks{E-mail: 
yoonbai@cosmos.skku.ac.kr}, and 
Phillial Oh${}^{{\rm a,b}}$\thanks{E-mail: ploh@ctpa03.mit.edu}}
\address{${}^{{\rm a}}$Department of Physics and Institute of Basic Research,
Sungkyunkwan University,\\ Suwon 440-746, Korea\\
${}^{{\rm b}}$Center for Theoretical Physics and Department of Physics,\\
Massachusetts Institute of Technology,
Cambridge, MA 02139, U.S.A.}

\maketitle
\begin{abstract}
We compute the entropy of systems of quantum particles satisfying the 
fractional exclusion statistics in the space-time of  2+1 dimensional black 
hole by using the brick-wall method. We show that the entropy of each 
effective quantum field theory with a Planck scale ultraviolet cutoff 
obeys the area law, irrespective of the angular momentum of the black hole 
and the statistics interpolating between Bose-Einstein and Fermi-Dirac 
statistics.
\end{abstract}

\vspace{5mm}

\indent\indent\hspace{4mm}  Keywords: BTZ black hole, Entropy, Exclusion 
statistics


\newpage

\section{Introduction}
Since the discovery of area law for black hole entropy
by Bekenstein~\cite{bekenstein} and Hawking~\cite{hawking},
considerable efforts have been made in order to
derive the thermodynamic properties in the contexts of both
Euclidean path-integral formulation~\cite{gibbons} and
microcanonical functional integral formulation~\cite{brown2}.
During the last decade, the statistical origin of the entropy 
of a black hole  has been discussed in connection with the information
approach~\cite{bekenstein,zurek} and with the entanglement
entropy~\cite{thooft}.
It is widely believed that the entropy of  quantum 
field in the black hole background satisfies the area law. 
One of the authors recently showed in Ref.~\cite{hckim} that 
the entropy of a quantum field with a Planck scale ultraviolet (UV) cutoff is 
proportional to the area of the black hole event horizon by
using the brick wall method and the microcanonical ensemble approach.

In the 80's, the 2+1 dimensional anti-de Sitter gravity 
attracted considerable attention 
because of its Chern-Simons gauge formulation~\cite{AT}. 
In the early 90's, Bar\~{n}ados, Teitelboim, and Zarnelli (BTZ)
reported 2+1 dimensional black hole solutions
in anti-de Sitter spacetime~\cite{BTZ}.
Extensive studies on the thermodynamics of 
BTZ black holes have  followed~\cite{Ther}.  
Recently, systematic counting of states for the
BTZ black hole was proposed, and the area law  was recovered
in the framework of statistical 
mechanics with some assumptions~\cite{Car}. 
Another intriguing feature in 2+1 dimensional
physics is  spin-statistics relation, that is,
anyons~\cite{Wil} or exclusion 
statistics~\cite{HW}. Therefore, a natural question
in 2+1 dimensional black hole physics  is its possible connection
with statistics. 

More specifically, we  ask whether
the area law is satisfied for the entropy of the quantum  
particles satisfying the fractional exclusion statistics
in the background of BTZ black hole.
An immediate obstacle is that we do not have a specific
quantum field theory example of elementary excitations satisfying exclusion 
statistics. Instead we assume the existence of such free field, so-called 
{\it exclusons}
in the background of the BTZ black 
holes, and simply impose a mass shell condition (Klein-Gordon equation).
Since ordinary particle physics based on a local quantum field theory fails 
near the horizon of a black hole~\cite{Hoo,LPS}, the valid
range of our effective field theory is restricted beyond
Planck scale UV cutoff.
In this paper we find that the area law is satisfied for each
quantum statistics irrespective of the statistical parameter 
connecting boson and fermion.

This paper is organized as follows.
In Sec.~II, we calculate the phase volume of one particle outside the  
BTZ black hole by using the brick-wall method,  and then consider
many particle systems with the energy and the particle number
fixed. Here
we examine the statistical behavior of the system which follows the
exclusion statistics.
Next we remove the constraint on the particle number and
take an ensemble sum over all particle number states
for the quantum statistics.  We conclude in Sec.~III with a few
discussions.

\section{Black Hole Entropy of Exclusons}
Let us begin this section with a brief recapitulation of BTZ black hole
solutions. 
The line element of  BTZ black hole as a vacuum solution is described by
\begin{eqnarray}
ds^2&=&g_{\alpha \beta}dx^\alpha dx^\beta\nonumber\\
&=& -\Phi^2 dt^2+ \Phi^{-2} dr^2 + r^2(N^\phi dt + d\phi)^2 ,
\label{btz}
\end{eqnarray}
where $\Phi^2(r) = -M + r^2/l^2 + J^2/4 r^2$, 
and $N^\phi(r) = - J/2r^2$~\cite{BTZ}.
Here $M$ and $J$ denote mass and angular momentum per unit mass
of the BTZ black hole respectively, and they satisfy $M>0$, 
and $|J| \leq Ml$. From Eq.~(\ref{btz}), the outer horizon of the 
black hole is present at
\begin{eqnarray}
r_+ =l \left[ \frac{M}{2} \left\{ 1+ \left[1- \left( \frac{J}{Ml}
\right)^2 \right]^{1/2} \right\} \right].
\end{eqnarray}
In the rotating frame with a constant azimuthal velocity $\Omega_H$ which
is the angular velocity of the horizon,
the metric (\ref{btz}) becomes
\begin{eqnarray}
\label{metr}
ds^2 = g_{tt}' dt'^2 + 2 g_{\phi \phi}\left( \Omega_H+
\frac{g_{t\phi}}{g_{\phi \phi} }
	\right) dt' d\phi' + g_{\phi \phi} d\phi'^2+ g_{rr} dr^2,
\end{eqnarray}
where $g_{tt}'= - \Phi^2 + r^2(N^\phi + \Omega_H)^2$.
Suppose that there is a quantum mechanical particle of mass $m$ outside the
black hole horizon. If it carries energy $E$
and momentum $(p_r , p_{\phi})$, it satisfies
the following dispersion relation:
\begin{eqnarray} \label{compact} 
\frac{-g_{tt}'}{-D} \left(p_\phi + \frac{g_{t \phi}+ \Omega_Hg_{\phi
\phi}}{g_{tt}'}   E \right)^2 +  \frac{p_r^2}{g_{rr}}
       = \frac{ E^2}{-g_{tt}'} - m^2 ,
\end{eqnarray}
where $-D= (g_{t\phi})^2- g_{tt} g_{t\phi}= r^2 \Phi^2$.

Let us consider $N$ quantum mechanical particles satisfying
Eq.~(\ref{compact}) in a box bounded by two concentric cylinders with radii
$r_+ +h$ and $L$. The nature of the particles' spin is assumed to appear in
their mutual statistics, and these $N$ particles satisfy 
exclusion statistics.
In this system of $N$ exclusons, energy levels $E_{i}$ and 
occupation numbers $n_{i}$ are specified by
an index $i$ so that the total energy $E$
\footnote{Since the spacetime
structure of the BTZ black hole is that of
a (2+1)-D anti de Sitter spacetime, it is not asymptotically flat.
Therefore, though the identification of $E$ as the particle energy is not
exact, we will attach the name ``energy'' to this constant of motion of
the timelike Killing vector.}
and the particle number $N$
satisfy the conditions:
\begin{eqnarray} \label{EN:i}
E= \sum_i E_i n_i ,~~~ N= \sum_i n_i .
\end{eqnarray}
Then the number of accessible states of this system is given by the sum of
the number of states of the system corresponding to the set of occupation
numbers $\{n_i \}$, i.e.,
\begin{eqnarray} \label{gN}
g_N(E) = \sum_{\{n_i\}} W(\{n_i\}) .
\end{eqnarray}
For the fractional exclusion statistics, $W(\{n_i\})$ is given by \cite{HW}
\begin{eqnarray}
W(\{n_i\})  =\prod_{i=1}\frac{[g_1(E_i)+(n_i-1)(1-\alpha)]! }{n_i![
                g_1(E_i)- \alpha n_i - (1-\alpha)]!},
\end{eqnarray}
where the statistics parameter $\alpha$ interpolates 
between boson$(\alpha=0)$ and fermion$(\alpha=1)$. 

If the governing relativistic quantum mechanical equation is local for each
excluson, then the energy-momentum dispersion relation (\ref{compact}) implies
that the wave function of each particle should satisfy the Klein-Gordon
equation irrespective of its statistics:
\begin{eqnarray} \label{Psi:0} 
(\nabla_\mu \nabla^\mu - m^2 ) \Psi =0   .
\end{eqnarray}
In the context of semiclassical approximation, the WKB solution has the
following form;
\begin{eqnarray} \label{Psi} 
\Psi = e^{-i  E t + i p_\phi \phi + i R(r) },
\end{eqnarray}
where $R(r)$ is obtained from $p_r = \partial R/\partial r$.
In a box normalization with an appropriate boundary condition
as in the brick wall method~\cite{thooft} with its  location at 
$r=r_+ + h$, a discrete momentum
eigenvalue corresponds to one quantum state per unit volume.
Then the sum over the quantum states can be rewritten as the
integral over the phase space. 
The proper distance $\epsilon$ from the horizon to the box is given
by
\begin{eqnarray} \label{eps}
\epsilon = \int_{r_+}^{r_++ h} dr \sqrt{g_{rr}}=\frac{2 \sqrt{h}}{
        \sqrt{\frac{\partial}{\partial r}{\Phi^2}|_{r=r_+}} },
\end{eqnarray}
and the area of the inner brick wall $A$ is given by
\begin{eqnarray}\label{area}
A=2\pi r_{+}.
\end{eqnarray}
Note that when   the inner radius of the box 
starts at $r=r_+$, i.e., $h\rightarrow 0$,  the inverse of
proper distance $\epsilon$ diverges and so do the many physical 
quantities (see below).
Since this boundary condition is a natural choice in the black 
hole geometry, a UV cutoff is unavoidable
in our local quantum field theory, and we  
shall relate  $\epsilon$ in Eq.~(\ref{eps}) with
 the Planck scale~\cite{Hoo,LPS}.

First of all let us compute the entropy of one particle system, and then
extend it to  that of $N$ particles. The phase volume of a classical particle
with fixed energy $E$ in the frame (\ref{metr}) is the volume of a
hypersurface satisfying $ H(p,x) = E$, that is, $ g_1(E) = \int d^2
p d^2 x \delta( E- H(p,x)) $, and is obtained by $\partial
\Gamma/\partial E$, where $\Gamma(E) =  \int d^2 p d^2 x \theta( E- H(p,x)) $.
By use of the dispersion relation (\ref{compact}), $\Gamma(E)$ is calculated
to be
\begin{eqnarray}\label{gamm}
\Gamma(E) =  \pi \int_{box} d^2x \sqrt{ \frac{-g_{rr} D}{-g_{tt}'}}
                \left( \frac{E^2}{-g_{tt}'} - m^2 \right) .
\end{eqnarray}
We integrate out the coordinate space integrals in Eq.~(\ref{gamm}), and
express it in terms of the distance $\epsilon$ (\ref{eps}) and the area
$A$ (\ref{area}). 
In the limit of $h \rightarrow 0$, these quantities are divergent;
\begin{eqnarray}
\Gamma(E) &\sim& \frac{4E^2}{\left(\left.\frac{\partial}{
        \partial r}\Phi^2 \right|_{r=r_+} \right)^2 }
        \frac{2 \pi r_+}{\epsilon}
        = \frac{E^2}{\kappa^2} \frac{A}{\epsilon},
\label{Gam}\\
g_1(E) &\sim& \frac{2 E}{\kappa^2} \frac{A}{\epsilon} ,
\label{g:A}
\end{eqnarray}
where $\kappa = \frac{\partial}{\partial r}{\Phi^2}/2|_{r=r_+}$ is the
surface gravity at the horizon~\cite{pad}.
Here let us recall that the model of our interest is an effective
local quantum field theory, and it is valid only with a UV cutoff in the BTZ
black hole background. 
Inserting Eq.~(\ref{g:A}) into the definition of the entropy of a particle, 
we get 
\begin{eqnarray}
S_{1} \equiv  \ln g_{1}(E) = \ln \frac{2 E}{\kappa^2} \frac{A}{\epsilon}.
\end{eqnarray}
Note that the obtained entropy is not seemed to be proportional to the area 
$A$. Since the one particle system does not
constitute a thermodynamical system and statistics of particles is not
necessary, the area law needs not to be preserved in this case.

Now let us  extend our calculation to many particle system.
The number of accessible states $g_{N}(E)$ for the system of total energy $E$
and total number of particles $N$ is estimated by
the maximal entropy principle when $N$ is sufficiently large. The value of
$g_N(E)$  can be replaced by the maximal value of $W(\{\bar{n}_i\})$,
where $\{\bar{n}_i \}$ is a set of occupation numbers that maximize
$W(\{n_i\})$ subject to Eq.~(\ref{EN:i}):
\begin{eqnarray} \label{n:zbeta}
\bar{n}_i = \frac{g_1(E_i)}{\omega(e^{\beta (E_i - \mu)}) +
\alpha} ,
\label{nbar}
\end{eqnarray}
where the temperature $1/\beta$ and chemical potential $\mu$ are determined as
functions of the total energy $E$ and number of particles $N$ in
Eq.~(\ref{EN:i}).
The function $\omega(\zeta)$ satisfies the functional equation
\begin{eqnarray}\label{ome1}
\omega(\zeta)^\alpha [1+ \omega(\zeta)]^{1-\alpha} =
		 \zeta \equiv e^{\beta(E-\mu)},
\end{eqnarray}
or its equivalent one which is obtained by differentiation 
\begin{eqnarray}\label{ome2}
\frac{\omega'(\zeta)}{\omega(\zeta)[\omega(\zeta)+1]}=  
\frac{1}{\zeta [\alpha +
	\omega(\zeta) ]}.
\end{eqnarray}
Note that $\omega(\zeta)=\zeta-1$ for bosons ($\alpha =0$) and
$\omega(\zeta)=\zeta$ for fermions ($\alpha=1$).

By making use of  the Stirling's formula, the entropy defined by
$S_N = \ln W(\{ \bar{n}_i \})$ becomes
\begin{eqnarray} \label{S}
S_{N} &=& \sum_{i=1} \left[ \bar{n}_i \ln
\left\{ 1+\frac{g_1(E_i)-\alpha \bar{n}_i
	-(1-\alpha)}{\bar{n}_i} \right\} \right. \\
  &&+\left. \left\{g_1(E_i)-\alpha \bar{n}_i -(1-\alpha) \right\}
	\ln\left\{  1+ \frac{\bar{n}_i}{g_1(E_i)- 
\alpha \bar{n}_i - (1-\alpha)}  
\right\}	\right].  \nonumber
\end{eqnarray}
Again a change from the summation to an integral is applied to Eq.~(\ref{S})
and some rearrangement of the terms in Eq.~(\ref{S}) with the aid of
Eqs.~(\ref{ome1}) and (\ref{ome2}) yields
\begin{eqnarray} \label{S:QM}
S_{N} = \frac{2A}{ (\kappa\beta)^2\epsilon} \left[ 3 f_{\alpha, 3}(\mu \beta)
		- \mu \beta f_{\alpha, 2}(\mu \beta)
	  \right] ,
\end{eqnarray}
where
\begin{eqnarray}\label{f1}
f_{\alpha, n}(\mu \beta)&=& \frac{1}{(n-1)!} \int_0^\infty dx
	\frac{x^{n-1}}{\omega(e^{x-\mu\beta})+ \alpha} .
\end{eqnarray}
Note that, for $n \geq 2$, $f_{\alpha,n}$ satisfies the following recursion  
relation:
\begin{eqnarray}
\label{f:eq1}
\frac{\partial}{\partial(\mu\beta)} f_{\alpha,n} = f_{\alpha,n-1}. 
\end{eqnarray}
Furthermore, $f_{\alpha,n}(\mu\beta)$ is monotonic decreasing function
of $\alpha$ because
\begin{eqnarray}
\label{f:alpha}
\frac{\partial}{\partial\alpha}[\alpha+\omega(\mu\beta)]= 
	1+\frac{\omega (1+ \omega)}{\omega + \alpha} 
	\ln\left[\omega(1+\omega) 
	\right] \geq 0 \mbox{ for } \omega > 0, ~ 1 \geq \alpha \geq 0.
\end{eqnarray}
The energy and the number of particles of the system given by the constraint 
equations (\ref{EN:i}) are rewritten in a form of integrals:
\begin{eqnarray}\label{constraint1}
E= \frac{2A}{(\kappa \beta)^2 \epsilon}~
	\frac{2}{\beta} f_{\alpha,3}(\mu \beta )  , ~~~
N-n_0= \frac{2A}{(\kappa\beta)^2\epsilon} f_{\alpha,2}(\mu \beta ) .
\end{eqnarray}
Here $n_{0}$ stands for the number of ground state particles, i.e., $n_0=
1/(\omega(e^{-\beta \mu})+\alpha)$ from Eq.~(\ref{n:zbeta}), and this term
should be taken into account for the case of bosons exclusively.
The entropy (\ref{S}) can be expressed in terms of the energy $E$ and the 
number of particles $N$ in Eq.~(\ref{constraint1}) as
\begin{eqnarray}
S_N=\beta\left[\frac{3}{2}E-\mu(N-n_{0})\right].
\end{eqnarray}
By use of Eq. (\ref{constraint1}), we have
\begin{eqnarray}
\left. \frac{\partial}{\partial E} S_{N}\right|_{N} = \frac{1}{T}=\beta , ~
\left. \frac{\partial}{\partial N}S_{N}\right|_{E} = -\mu \beta .
\end{eqnarray}

Since any local quantum field theory of point particles in black hole
background is believed to include  divergences~\cite{Hoo,LPS}, 
we defined our theory as an effective quantum field 
theory with an
external UV cutoff. The obtained results in Eqs.~(\ref{S}) and
(\ref{constraint1}) also contain the 
cutoff $1/\epsilon$ explicitly so that we
have to adjust the cutoff. We are interested in the question whether or not
the dependence on the 
statistics parameter $\alpha$ appears in the expression of the
Bekenstein-Hawking entropy of the black hole ($4\pi r_+$). Therefore, an
appropriate choice of the UV cutoff is to use the Planck scale which is the
only relevant scale in our theory;
\begin{eqnarray}\label{epsilon}
4 \pi^2 \epsilon = 3 f_{0,3}(0) l_P .
\end{eqnarray}
With this cutoff the energy $E$ and the number of particles $N$ in
Eq.~(\ref{constraint1}) become
\begin{eqnarray}\label{EN}
E = \frac{4}{3 \beta_H} \frac{f_{\alpha,3}(\mu\beta_H)}{f_{0,3}(0)}
	\frac{2 \pi r_+}{l_P}, ~
N -n_0= \frac{2 f_{\alpha,2}(\mu\beta_H) }{3 f_{0,3}(0)} \frac{2 \pi r_+}{l_P},
\end{eqnarray}
where $\beta_H= 2 \pi/\kappa$ is on-shell temperature.
We attempt an expansion of the entropy (\ref{S}) for small chemical potential
with $E$ fixed:
\begin{eqnarray}
S_N = \frac{4\pi r_+}{l_P} \frac{f_{\alpha,3}(0)}{f_{0,3}(0)}
	\left[  1-\frac{1}{2}
        \left( \frac{f_{\alpha,1}(0) f_{\alpha,3}(0)}{f_{\alpha,2}^{~2}(0)}-2
        \right)
        \left( \frac{f_{\alpha,2}(0)}{3f_{\alpha,3}(0)} \right)^2
         (\mu \beta_H)^2 + \cdots \right].
\end{eqnarray}
The area law is satisfied only when the chemical potential $\mu$ vanishes.

Now we remove the conservation of total particle number $N$ of the system and
allow the transition between states of different particle numbers 
in order to describe the many body theory of exclusons.
Therefore, when we count the accessible states in the microcanonical
ensemble for a quantum field, we have to take into account all possible 
particle number states with the total energy being fixed:
\begin{eqnarray}
\label{constraint:QF} 
E= \sum_i n_i E_i .
\end{eqnarray}
The number of states $g(E)$ can be obtained by summing $g(N)$ over 
all possible $N$:
\begin{eqnarray}
\label{gE}
g(E)= \sum_{N=0}^{\infty}  g_N(E) .
\end{eqnarray}
This summation may be approximated appropriately by the 
peak value of ${\bar N}$
which maximizes $S=\ln g_N(E)$ by the following equation
\begin{eqnarray}
\label{vareq}
dS_N |_E &=& \mu \beta d(N-n_0) =0 . 
\end{eqnarray}  
The unique solution to this equation (\ref{vareq}) is $\mu=0$ and 
$\displaystyle {\bar N}= n_0+ \frac{2f_{\alpha,2}(0)}{3f_{0,3}(0)} 
\frac{2\pi r_+}{l_P}$ from Eq.~(\ref{EN}). Now, the total energy
(\ref{constraint:QF}) is replaced by the expression in Eq.~(\ref{constraint1})
with zero on-shell chemical potential $\mu_{H}=0$, and then the entropy 
$S= \ln g(E)$  
of the system with the constraint (\ref{constraint:QF}) is given by
\begin{eqnarray}
S\approx S_{\bar{N}} = \frac{6A}{(\kappa \beta)^2\epsilon} f_{\alpha,3}(0) 
	= \frac{3}{2} \beta\left. E\right|_{\mu=0}.
\end{eqnarray}
Under the on-shell temperature $\beta_{H}$ and the UV cutoff (\ref{epsilon}),  
the maximum entropy reproduces the area law:
\begin{eqnarray} \label{S:QF}
S = \frac{3}{2}\beta_H \left. E\right|_{\mu=0} =
\frac{f_{\alpha,3}(0)}{f_{0,3}(0)} \frac{4 \pi r_+}{l_P}.
\end{eqnarray}
A remark on the $\alpha$-dependence should be placed: the coefficient
$f_{\alpha,3}(0)/f_{0,3}(0)$ is a monotonic decreasing function of 
$\alpha$, and reproduces successfully the values at both ends, i.e.,
1 for bosons $(\alpha=0)$ and $3/4$ for
fermions $(\alpha=1)$~\cite{HW}.

\section{Conclusion}

In this paper, we have computed the entropy of quantum particles
obeying the fractional exclusion statistics in the background of 
the BTZ black holes with or without angular momentum. 
The only assumption we have made was the existence of
exclusons whose dispersion relation is consistent with the
locality of the corresponding field. 
It has been shown for quantum statistical systems 
that the area law is satisfied irrespective of the species of particles,
and the spin dependence comes through an over-all proportionality
constant. 

Since our calculation was based on the  
brick-wall method with   
external UV cutoff, a genuine defect of
this method also appeared in our formula: the entropy contained an {\it ad
hoc} cutoff introduced in order to control the divergence from infinite phase
volume around the event horizon.  We regularized it by setting this UV cutoff
to be a specific value of Planck length scale which must be the only natural
cutoff scale in our model.

A final comment is in order: Since the fractional exclusion statistics
holds in any space-time dimension~\cite{HW}, the area law may 
also be derived for the system of exclusons in both 1+1 and 3+1
dimensions.

\acknowledgments{This work was supported by the Ministry of Education
(BSRI/97-1419), the KOSEF (Grant No. 95-0702-04-01-3 and 
through CTP, SNU), and Faculty Research Fund, Sungkyunkwan University, 1997.}

\end{document}